\documentclass[%
 aip,
 amsmath,amssymb,
 reprint,%
nolongbibliography, 12pt]{revtex4-1}
\usepackage[left=2cm, right=2cm, top=2.5cm, bottom=2.5cm]{geometry}
\usepackage{etoolbox}
\makeatletter
\patchcmd{\thebibliography}{\section*{\refname}}{}{}{}
\patchcmd{\thebibliography}{\section*{References}}{}{}{}
\makeatother
\usepackage{graphicx} 
\usepackage[section]{placeins}
\usepackage{dcolumn}
\usepackage{bm}
\usepackage[utf8]{inputenc}
\usepackage[T1]{fontenc}
\usepackage{mathptmx}
\usepackage{etoolbox}
\usepackage{setspace}
\usepackage{xcolor}
\usepackage{bm}

\makeatother
\begin{document}

\definecolor{Red}{rgb}{1,0,0}
\definecolor{Black}{rgb}{0,0,0}
\newcommand{\revision}[1]{\textcolor{Black}{{#1}}}
\title[]{Uniaxial strain-driven ferroelastic domain control in LaAlO$\bm{_3}$}

\author{Matthias Roeper}

\author{Robin Buschbeck}
\affiliation{Institut für Angewandte Physik, TU Dresden, Nöthnitzer Straße 61, 01187 Dresden, Germany\looseness=-1}%

\author{Jakob Wetzel}
\affiliation{Institut für Angewandte Physik, TU Dresden, Nöthnitzer Straße 61, 01187 Dresden, Germany\looseness=-1}%
\affiliation{
Würzburg-Dresden Cluster of Excellence (EXC2147) ctd.qmat – Complexity, Topology, and
Dynamics in Quantum Matter, Dresden, Germany\looseness=-1
}

\author{Tobias Ritschel}
\affiliation{
Institute of Solid State and Materials Physics
Technische Universität Dresden
01069 Dresden, Germany\looseness=-1
}

\author{Anna-Lena Hofmann}
\affiliation{Institut für Angewandte Physik, TU Dresden, Nöthnitzer Straße 61, 01187 Dresden, Germany\looseness=-1}

\author{Vladyslav Kovtunovych}
\affiliation{Institut für Angewandte Physik, TU Dresden, Nöthnitzer Straße 61, 01187 Dresden, Germany\looseness=-1}

\author{Mike N. Pionteck}
\affiliation{
Institut für Theoretische Physik and Center for Materials Research (LaMa/ZfM),
Justus-Liebig-Universität Gießen, 35392 Gießen, Germany\looseness=-1
}

\author{Javier Taboada-Gutiérrez}
\affiliation{
University of Geneva, Department of Quantum Matter Physics (DQMP), Geneva 1211,
Switzerland\looseness=-1
}

\author{Alexey B. Kuzmenko}
\affiliation{
University of Geneva, Department of Quantum Matter Physics (DQMP), Geneva 1211,
Switzerland\looseness=-1
}

\author{Martina Basini}
\affiliation{
Physics Department,
ETH Zurich,
Zürich 8093, Switzerland\looseness=-1
}

\author{Vivek Unikandanunni}
\affiliation{
Institute of Applied Physics,
University of Bern,
Bern 3012, Switzerland\looseness=-1
}
\author{Iuliia Kiseleva}
\affiliation{Institut für Angewandte Physik, TU Dresden, Nöthnitzer Straße 61, 01187 Dresden, Germany\looseness=-1}

\author{Jochen Geck}
\affiliation{
Institute of Solid State and Materials Physics
Technische Universität Dresden
01069 Dresden, Germany\looseness=-1
}

\author{Susanne C. Kehr}
\affiliation{Institut für Angewandte Physik, TU Dresden, Nöthnitzer Straße 61, 01187 Dresden, Germany\looseness=-1}

\author{Maximilian Lederer}
\affiliation{Fraunhofer Institute for Photonic Microsystems IPMS
Center Nanoelectronic Technologies (CNT)
01109 Dresden, Germany\looseness=-1}

\author{Simone Sanna}
\affiliation{
Institut für Theoretische Physik and Center for Materials Research (LaMa/ZfM),
Justus-Liebig-Universität Gießen, 35392 Gießen, Germany\looseness=-1
}

\author{Lukas M. Eng}
\affiliation{Institut für Angewandte Physik, TU Dresden, Nöthnitzer Straße 61, 01187 Dresden, Germany\looseness=-1}
\affiliation{
Würzburg-Dresden Cluster of Excellence (EXC2147) ctd.qmat – Complexity, Topology, and
Dynamics in Quantum Matter, Dresden, Germany\looseness=-1
}%

\author{Samuel D. Seddon}
\affiliation{Institut für Angewandte Physik, TU Dresden, Nöthnitzer Straße 61, 01187 Dresden, Germany\looseness=-1}%
\altaffiliation{samuel.seddon@tu-dresden.de}

\date{\today}
\onecolumngrid
\maketitle
\onecolumngrid

\doublespacing

\textbf{Multiferroic domain walls in functional oxides exhibit properties distinct from the bulk and are increasingly exploited as active elements in nanoelectronic and photonic devices. Deterministic control of domain populations has typically remained limited to local control, or removal with temperature. Here we demonstrate continuous, reversible manipulation of the ferroelastic domain structure in single-crystal LaAlO$\bm{_3}$ using in-situ uniaxial strain. Combining atomic force microscopy, X-ray diffraction, and Raman spectroscopy with first-principles calculations we map the complete microscopic evolution of the twin domain population through the strain-driven transition from the rhombohedral $\bm{R\bar{3}c}$ ground state toward the predicted orthorhombic $\bm{Fmmm}$ phase. Applied strains below $\bm{0.5\%}$ produce pronounced surface flattening and large-scale domain reorganisation, establishing uniaxial strain as a technically accessible control parameter for ferroelastic domain engineering. These results open a route to active, real-time programming of domain architectures in LaAlO$\bm{_3}$-based heterostructures, with implications for strain-tunable superconducting interfaces, nanoscale phonon-polariton optics, and ultrafast lattice control.}

Ferroic materials possess switchable order parameters—electric polarization, magnetization, and strain—whose reversible control by conjugate fields underpins a broad spectrum of modern technologies\cite{Catalan_2012, Meier_2022}. The boundaries between domains, known as domain walls, have emerged as functional nanoscale objects whose properties differ markedly from the surrounding bulk, opening routes to memristive, synaptic, and other nanoelectronic device concepts\cite{Catalan_2012, Meier_2022, Nataf_2020}. Their high spatial mobility enables real-time tunability through external stimuli such as electric fields in ferroelectrics\cite{Das_Adhikary_2025} or mechanical stress in ferroelastics\cite{Lu_2012}. Ferroelasticity, the largest ferroic class\cite{Salje_2020}, is characterised by hysteretic switching between two or more crystallographic orientation states under mechanical stress. The spontaneous strain accompanying a ferroelastic phase transition gives rise to complex twin microstructures whose wall geometry and population distribution are intimately coupled to the local strain state\cite{Bueble1999, Bueble_1998, Guo_2019}.

Beyond their role as model systems for symmetry breaking, ferroelastic domains actively mediate the functional response of a wide class of materials: they govern the electromechanical output of piezoelectrics and multiferroics\cite{Manosa_2010}, control mechanical dissipation near structural phase transitions\cite{Harrison_2004, Harrison_2004b, Schranz_2012} with ferroelastic composites enabling extreme damping\cite{Lakes_2001} and serve as targets for nanoscale domain engineering\cite{Sood_2021, Nataf_2020}. Stabilised ferroelastic switching further enables non-volatile magnetoelectric operation relevant to high-density memories\cite{Baek_2010}, while in inorganic perovskites epitaxial stress at domain walls has been exploited to nucleate unusual two-dimensional phases with emergent chemical and magnetic order, an effect predicted to be generic to strained orthorhombic $ABO_3$ systems\cite{Farokhipoor_2014}. Controlled manipulation of ferroelastic domain populations therefore represents a central challenge, with direct implications for strain-tunable oxide heterostructures. Whilst some success has been achieved on the local scale \cite{Seidel_2009, Bea_2011, Peng_2025}, global stress/strain control has received relatively scant attention. As device miniaturisation drives ever-increasing stress in substrate–film systems, such control may enable active domain engineering from tunable superconducting interfaces to atomic-scale diffusion channels\cite{Salje_2023}.

Lanthanum aluminate (LaAlO$_3$; LAO), a ferroelastic perovskite oxide, serves as the canonical model system for ferroelastic domain physics\cite{P_Bouvier_2002, Harrison_2010, Puchberger_2017, Salje_2016}, exhibiting polar tweed textures that connect to broader questions of hierarchical pattern formation\cite{Salje_2016}. As a substrate, the LAO domain structure directly governs the properties of overlying thin films, including the critical temperature of superconductors\cite{Bellingeri_2012}, the magnetic anisotropy of ferromagnetic layers\cite{Hussain_2019}, and direction-dependent lattice distortions in BiFeO$_3$\cite{Hatt_2010, Yang_2019}. As a constituent layer of the two-dimensional electron gas at the LAO/SrTiO$_3$ (STO) heterointerface\cite{Ohtomo_2004}—which hosts conductivity, superconductivity, and ferromagnetism\cite{Lee_2013, Bert_2011, Hwang_2012}, gate-tunable through a superconductor–insulator quantum critical point\cite{Caviglia_2008} without the chemical disorder of conventional doping\cite{Hwang_2012}—LAO is of particular relevance, as these interface properties are sensitive to the strain state of the LAO layer. Recent demonstrations of resonantly driven infrared-active\cite{Gollwitzer_2025} and parametrically excited Raman-active\cite{Basini_2026} phonon modes further point to ultrafast lattice-control capabilities, while LAO domain walls offer a possible platform for nanoscale light confinement via phonon-polariton canalization\cite{Wetzel_2026}. Below a structural phase-transition temperature of $544~^\circ$C, LAO condenses from the undistorted cubic $Pm\bar{3}m$ phase via a soft-mode-driven antiferrodistortive transition\cite{Guennou_2011} into the rhombohedral $R\bar{3}c$ space group\cite{Bueble1999}, supporting a rich variety of twin domain patterns\cite{Bueble1999, Salje_2016, Bueble_1998} with up to four symmetry-allowed domain states on a (001) surface\cite{Bueble1999}. Previous efforts to manipulate this domain structure have relied primarily on temperature\cite{Hayward_2005} and hydrostatic pressure, applied to both single crystals\cite{Guennou_2011} and powders at pressures up to 63~GPa\cite{P_Bouvier_2002}, as well as on dynamic mechanical experiments\cite{Harrison_2004, Harrison_2004b}. First-principles calculations predict that the orthorhombic $Fmmm$ phase should become stable under comparatively modest uniaxial stress below $1~$GPa\cite{Hatt_2010}, suggesting uniaxial strain as a highly accessible control parameter. However, early uniaxial studies were limited to macroscopic characterisation of sintered polycrystalline material, where voids and grain boundaries precluded microscopic insight into the intrinsic domain response\cite{Islam_2017, Araki_2016}. Here we address this gap directly, combining scanning probe microscopy, X-ray diffraction, and Raman spectroscopy with first-principles calculations to establish a comprehensive microscopic picture of the strain-driven ferroelastic phase transition in single-crystal LAO.

\begin{figure*}[t]
    \centering
    \includegraphics[width=1\linewidth]{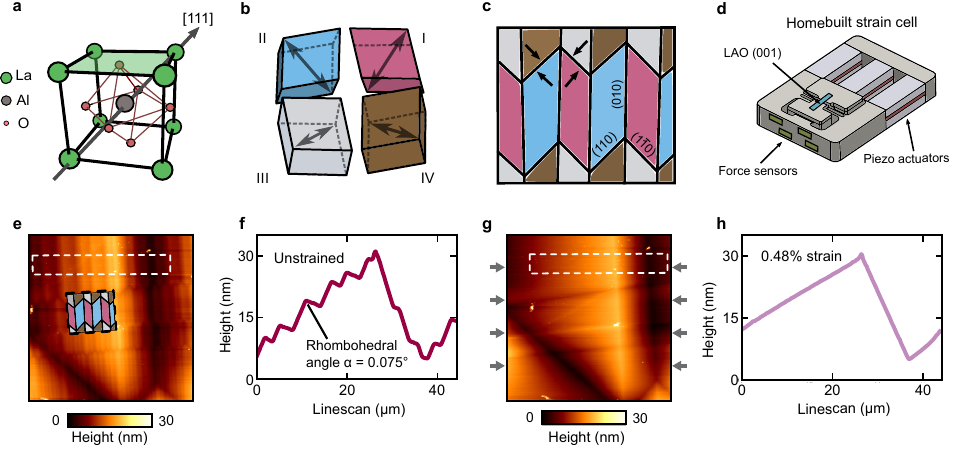}
    \caption{\textbf{LAO lattice structure, domain formation and strain response.} | \textbf{a} Pseudo-cubic LAO lattice with oxygen octahedra (red) slightly rotated about the [111] axis. \textbf{b} Four symmetry-allowed domain orientations on the (001) surface with characteristic strain distortions. \textbf{c} Chevron twin pattern on the (001) LAO surface (transparent green \revision{surface in \textbf{a}}) comprising four domain types separated by walls along the [010] and [110] directions. \textbf{d} Schematic of the homebuilt piezo-driven strain cell. \textbf{e} $(50\times50)~\mu\text{m}^2$ AFM height scan revealing ferroelastic micro-domains (with a few micrometer periodicity) superimposed by non-ferroelastic structures (few tens of micrometer periodicity). \textbf{f} Vertically averaged line scans (white dashed region in \revision{\textbf{e}}) showing inclined surface normals forming a stepped structure. \textbf{g} AFM scan under $0.48\%$ applied compressive strain (strain axis indicated by grey arrows) showing a strongly reduced surface substructure. \textbf{h} AFM height line scan under strain demonstrating surface flattening relative to \textbf{f}.
    }
    \label{fig:placeholder}
\end{figure*}

\section{Results and Discussion}

\textbf{Surface Inclination Control}\\
The pseudo-cubic perovskite representation of the LAO lattice (Fig.~1a) highlights the characteristic rotation of the oxygen octahedra about the [111] axis, which drives the rhombohedral distortion from the ideal cubic perovskite structure. Below the ferroelastic phase transition, this symmetry breaking reduces the cubic point group to rhombohedral symmetry, while on the (001) surface four symmetry-allowed domain variants emerge (see Fig.~1b), related to one another by the lost cubic symmetry operations. Each variant is characterised by a distinct in-plane strain distortion whose orientation reflects the projection of the rhombohedral distortion axis onto the (001) plane. The four variants are not independent: they are grouped into two pairs of twin domains, and the interfaces between them are constrained by compatibility conditions \cite{Sapriel_1975} to lie along the [010] and [110] directions. The collective arrangement of these variants across the surface gives rise to the characteristic chevron twin pattern (Fig.~1c), in which domain walls form continuous zig-zag lines alternating between the two allowed wall orientations and straight lines  along the [010] direction. This highly regular pattern is a direct consequence of the ferroelastic nature of the transition and the strict geometric constraints imposed on compatible domain wall orientations in a system with fourfold surface symmetry. Uniaxial strain\revision{, in this work,} is applied using the homebuilt piezo-driven strain cell\cite{Pionteck_2025, Singh_2023, Singh_2022} (Fig.~1d), in which stress is transmitted to the sample held between two plates and monitored by calibrated strain gauges. 

The AFM topography image acquired under ambient conditions (Fig.~1e) resolves this domain structure in detail. Ferroelastic domains with periodicities of a few $\mu$m in width are clearly present across the surface. Importantly, these are superimposed by larger surface undulations with periodicities of few tens of $\mu$m, which show self-similarity to the walls present in the microstructure, but are a real surface topography rather than ferroelastic domain structure. This scale invariant effect present in ferroelastics is already well established, and is the result of repeated polishing and annealing\cite{Bueble_1998}. Vertically averaged line scans extracted from the white dashed region (Fig.~1f) confirm that neighbouring ferroelastic domains present distinctly inclined surface normals, forming a stepped profile whose amplitude directly reflects the magnitude of the rhombohedral distortion. This careful distinction between these two classes of surface structure\revision{, i.e. between actual ferroelastic domains and pure topographic features,} is essential for the unambiguous interpretation of strain-dependent scanning probe data presented below.


 Upon the application of 0.7~GPa uniaxial compressive \revision{stress}, the surface morphology undergoes a dramatic transformation (Fig.~1g), with the ferroelastic domain structure visible under ambient conditions strongly suppressed, while the larger-scale non-ferroelastic topographic features persist as expected.   Consistent with this, the stepped surface profile observed in Fig.~1f flattens considerably under strain (Fig.~1h), reflecting a near-complete redistribution of the ferroelastic domain population. This behaviour is indicative of a strain-driven selective stabilisation of domain variants favourably aligned with the applied uniaxial strain direction, at the expense of unfavourably oriented variants. The fact that this redistribution occurs at moderate strain levels suggests that the energy barrier separating competing domain orientations in LAO, is small, rendering the ferroelastic domain structure highly susceptible to external mechanical perturbation. Closer examination of the two AFM images reveals the presence of some faint ferroelastic domains\revision{, extending across the whole sample horizontally (see Fig.~3a$_4$ / Fig.~S2b,} being still present under applied strain, implying that in fact, whilst reoriented, there are still some twin walls and therefore structural difference between horizontal pairs of domains. 

\textbf{Advanced structural analysis}\\
\begin{figure}[t]
    \centering
    \includegraphics[width=0.5\textwidth]{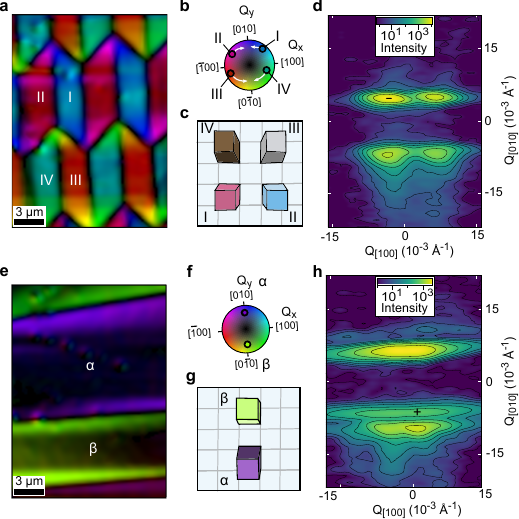}
    \caption{\textbf{Structural transition probed by AFM and XRD} | \textbf{a} Doubly differentiated AFM image (effectively mapping the surface normal) under ambient conditions where surface inclination direction and amplitude are colour-coded as indicated in \textbf{b}. Four domain types are present as expected from Fig.~1c, labelled I--IV. \textbf{b} Inclination directions (I--IV) extracted from the AFM scan, with the corresponding rhombohedral distortions shown exaggerated in \textbf{c}. \textbf{d} 3D-XRD measurement projected onto the $Q_{[100]}$--$Q_{[010]}$ plane (in-plane directions), showing four peaks consistent with the four domain types identified by AFM.  \textbf{e} Doubly differentiated AFM scan at the same position as \textbf{a} under applied stress of 0.7~GPa. Domains I and III vanish with increasing strain while domains II and IV extend over the entire scan area. \textbf{f} Under strain, the inclinations of the two remaining domain types \revision{(denoted as $\alpha$ and $\beta$)} are oriented along the [001] axis, as expected of two orthorhombic domains \textbf{g}. \textbf{h} 3D-XRD projection under maximal strain; the four peaks merge into two, as well as the emergence of a third peak \revision{(indicated with a black '+' right above the main peak)}.}
\end{figure}
In order to further analyse the remaining domain structure under applied strain, the AFM topography images were  differentiated along the two orthogonal in-plane directions to create a local map of surface normal. The doubly differentiated AFM image acquired under ambient conditions can be seen in  Fig.~2a,  with inclination direction and amplitude colour-coded for clarity (Fig.~2b). Four distinct domain types are resolved, consistent with the symmetry-allowed variants anticipated from Fig.~1c and labelled I--IV, whose corresponding rhombohedral distortions are illustrated in Fig.~2b. Here, the colour scale can be interpreted as the in-plane direction, with each surface normal pointing out of the page. The presence of all four variants (i.e. as in the schematic Fig.~2c) is independently confirmed by 3D-XRD reciprocal space mapping\revision{. I}maging the (002) specular Bragg reflection along all three Q-space directions allows for the projection of the splitted peak onto the $Q_{[100]}$--$Q_{[010]}$ plane (Fig.~2d), which reveals four well-separated sub-peaks whose positions are fully consistent with the four domain orientations identified by AFM. The combined real- and reciprocal-space characterisation thus establishes an unambiguous correspondence between the surface domain structure and the underlying bulk lattice distortions. From the angle between each stepped terraces (0.075$^\circ$ at room temperature) follows directly the rhombohedral crystallographic angle of $\alpha' = 90^{\circ}+\alpha$. This is in good agreement with the averaged surface tilt determined via the XRD measurements of $\Delta Q_\parallel \cdot Q_\perp^{-1} = 0.091^\circ$, and consistent with previous literature values such as Bueble et al. (0.096$^\circ$)\:\:\cite{Bueble_1998}. 

Upon application of uniaxial strain \revision{along the [100] axis}, the domain structure undergoes a clear and reproducible reconfiguration. The equivalent image acquired under strain (Fig.~2e) shows  domains I and III progressively vanish with increasing strain (also demonstrated for a larger area in Fig.~3 a$_1$-$a_5$), while domains II and IV expand to cover the entire scan area (hereby labelled $\alpha$ and $\beta$). The two surviving, transformed domain types retain inclinations oriented along the [001] axis (Fig.~2f), consistent with the expected orthorhombic distortion geometry of a two-variant ferroelastic state (Fig.~2g). Fig.~2f shows the effect of these surface tilts on the in-plane colour map, with a strong surface tilt still present between domains, as reflected in schematic Fig.~2g.
\begin{figure*}[htb!]
    \centering
    \includegraphics[width=1\linewidth]{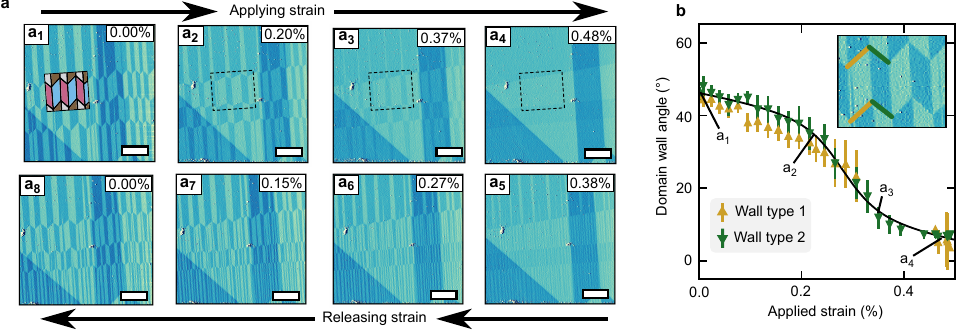}
    \caption{\textbf{In-plane twin wall angle rotation} | \textbf{a} Full strain cycle applying (\textbf{a1}--\textbf{a4}) and releasing (\textbf{a5}--\textbf{a8}) compressive strain\revision{, along the horizontal axis}, up to $0.48\%$ ($\sim$0.7~GPa stress) captured by AFM error signal scans (scale-bar 10~$\mu m$). \textbf{a1} Domain walls form zig-zag lines inclined at $45^\circ$ as expected from Fig.~1c. With increasing strain the angle decreases continuously until the walls flatten to $0^\circ$ (\textbf{a4}), accompanied by one domain type progressively consuming the surface area of the other. \textbf{a5}--\textbf{a8} The process is fully reversible: upon strain release the domain wall angle increases and the surface share of the receding domain type recovers. \textbf{b} In-plane domain wall angle with resprect to the (010) plane as a function of applied strain, extracted from the dashed region in \textbf{a1}--\textbf{a4}, analysed independently for the two domain wall orientations (\revision{yellow and green}), a line (black) is a guide to the eye. The domain wall angle evolves continuously with strain.}
\end{figure*}
This strain-driven symmetry reduction from a four-variant to a two-variant domain population is also  confirmed in reciprocal space: the 3D-XRD projection acquired at maximum applied strain (Fig.~2h) shows that the four peaks present under ambient conditions merge into two, directly reflecting the suppression of the unfavourably oriented domain variants. Given that a $(100)$ twin wall is strictly prohibited in this crystal setting, such a\revision{n} increase in symmetry \revision{despite} a lasting presence of distinct domains, is fully consistent with that of an orthorhombic $F\textit{mmm}$ structure. Indeed this was previously predicted by DFT calculations of uniaxial strain by Hatt and Spaldin\cite{Hatt_2010} resulting in a stable orthorhombic structure, though no discussion as to domains therein were addressed. 

The central result of this combined AFM and XRD analysis is that uniaxial strain drives a controlled, reversible transition from a four-variant to a two-variant ferroelastic domain state, and also a rhombohedral  $R\bar{3}c$ -- orthorhombic $Fmmm$ phase transition, with the domain redistribution manifesting simultaneously and consistently in both real and reciprocal space.

\textbf{Continuous twin wall rotation}

 The AFM error signal can be interpreted as a fast-scan axis derivative, and is therefore clearly correlated to topographic features.  AFM error signal scans acquired across a full strain cycle (Fig.~3a$_1$--a$_8$) provide a detailed real-space picture of the domain wall dynamics underlying the rhombohedral to orthorhombic phase transition already established. Applied strain values were acquired from in-plane lattice parameter shift in XRD measurements \revision{(see supplement F)}.  Under ambient conditions (Fig.~3a$_1$), domain walls form continuous zig-zag lines inclined at $\sim45^\circ$ to the strain axis, consistent with the chevron pattern expected from the fourfold domain symmetry in LAO\cite{Bueble_1998}. As uniaxial strain is applied, this angle decreases continuously and monotonically, with the domain walls rotating progressively toward alignment along the strain axis until they flatten to $0^\circ$ at maximum applied strain (Fig.~3a$_4$). This angular evolution is accompanied by a simultaneous redistribution of domain surface area: within each horizontal line of the scan, one domain type progressively expands at the expense of the other, until the minority variant is fully consumed at maximum strain. Together, the wall rotation and the surface area redistribution represent two complementary manifestations of the same underlying strain-driven domain repopulation process\revision{. B}oth processes are fully reversible: Upon release of the applied strain (Fig.~3a$_5$--a$_8$), the domain wall angle increases again and the surface share of the previously suppressed domain type recovers, with the surface returning to the ambient $45^\circ$ chevron configuration. As discussed, a small amount of hysteresis is seen, as even in zero field needle domains continue to propagate throughout the sample, however given sufficient relaxation \revision{($\tau$ = 4.8~h see supplement Fig.~S1)} Fig~3a$_8$ returns to the equivalent domain state as Fig.~3a$_1$.

To quantify the wall rotation, the in-plane domain wall angle was extracted as a function of applied strain independently for the two domain wall orientations  present on the surface, initially oriented along the ($110$) and ($\bar{1}10$) axes and denoted as wall type one and two (Fig.~3b, red and purple), within the region indicated by the dashed box in Fig.~3a$_1$--a$_4$. Both orientations exhibit a continuous, monotonic decrease with increasing strain. The continuity of this angular evolution, with no evidence of an abrupt transition or threshold behaviour typical of Barkhausen noise is somewhat unexpected, and nevertheless implies a smooth, second-order phase transition between the two crystal symmetries.

\textbf{\revision{Lifting} of phonon degeneracy}\\
\revision{LAO is an improper ferroelastic, its order parameter} is in fact not strain directly (though \revision{strain and the order parameter} are directionally coupled) bu\revision{t t}he octahedral tilt angle\cite{Howard_2004}. In order to probe the order parameter of this system most directly, this mode must be directly visualised. The first three Raman active modes are a $E_g$ mode, a symmetric non-degenerate $A_{1g}$ mode and a second higher energy degenerate $E_g$ mode. These are respectively characterised by oxygen octahedra rotation, La atom oscillation along the [111] axis, and La atom oscillation within a pseudo-cubic plane. A schematic of these three modes can be found in Fig.~4a. All three modes are clearly resolved in the ambient backward-scattered Raman spectrum in Fig.~4b, with peak positions in good agreement with the \revision{specifically performed} first-principles predictions, providing a firm spectroscopic baseline against which strain-induced shifts can be assessed.
\begin{figure*}[tb]
    \centering
    \includegraphics[width=1\linewidth]{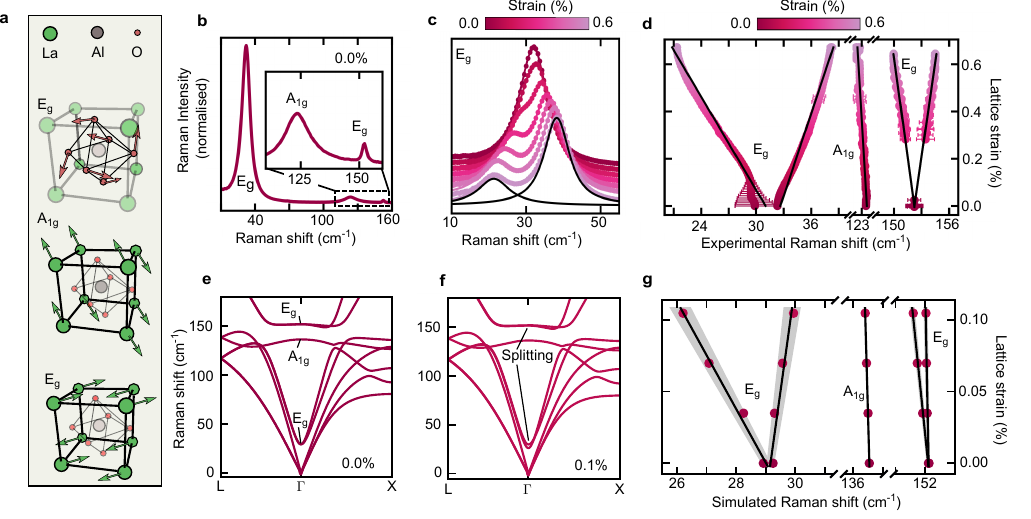}
    \caption{\textbf{Strain-dependent phonon behaviour of LAO} | \textbf{a} Three major modes in LAO corresponding to two $E_g$ modes and one $A_{1g}$ mode energetically between them. \textbf{b} Typical Raman spectrum of LAO (001) under ambient conditions showing three peaks. \textbf{c} Strain-dependent Raman spectra in the vicinity of the most intense $E_g$ mode displaying a strain-induced lifting of degenerac\revision{y (}data fitted with a pair of Voigt profiles in black\revision{)}. \textbf{d} $E_g$ peak splitting and $A_{1g}$ peak softening as a function of strain, extracted from peak fits and extracted values linearly fitted (black). \textbf{e} First-principles phonon dispersion of \revision{unstrained} LAO near the $\Gamma$-point. \textbf{f} Phonon dispersion under $0.1\%$ strain, showing lifting of $E_g$ degeneracy at the $\Gamma$-point. \textbf{g} Modal behaviour of calculations performed across several strain levels between $0\%$ and $0.1\%$ confirming the  behaviours for both the $E_g$ splitting and the $A_{1g}$ mode softening, as indicated by linear fits (black). 
    }
\end{figure*}
The strain dependence of the phonon spectrum is best displayed in the behaviour of the highest intensity $E_g$ mode (Fig.~4c). Under applied uniaxial strain, the single ambient $E_g$ peak progressively splits into two resolved components, directly evidencing a strain-driven lifting of the rotational degeneracy of the oxygen octahedra vibrations. This splitting is fully consistent with the symmetry reduction from a rhombohedral to a orthorhombic phase, with the application of uniaxial strain breaking the equivalence of the two $E_g$ sub-modes by selecting a preferred distortion axis, rendering them spectroscopically distinct. Systematic peak fitting across all strain levels (Fig.~4d) confirms that the $E_g$ splitting increases for both degenerate modes approximately linearly with applied strain, while the $A_{1g}$ mode exhibits a continuous softening, consistent with hydrostatic studies \cite{P_Bouvier_2002, Guennou_2011}. 

To further explore this behaviour, Density Functional Perturbation Theory (DFPT) calculation\revision{s a}llow the behaviour under uniaxial strain to be further explored. Here, the phonon band structure around the $\Gamma$-point reveal\revision{s} three lowest energy Raman-active phonon modes (Fig.~4e) as measured in Fig.~4b. By deforming the relaxed unit cell, based on the calculated compliance tensor, according to a pressure acting along the pseudo-cubic $[100]$ direction (the same direction as in the experiments), and therefore simulating the effect of applied stress, a visible splitting can be seen in the maximal simulated case (0.1\% strain, Fig.~4f) at the bands around the $\Gamma$-point. By plotting each mode as before, the relative behaviours as a function of intermediate values of strain \revision{reveal} a similar \revision{lift} of degeneracy of the $E_g$ modes. A broadly linear splitting is achieved as well as a phonon softening of the \revision{$A_{1g}$} mode. 


\section{Conclusion and Outlook}
The present work demonstrates that uniaxial strain provides a highly accessible and reversible handle for engineering the ferroelastic domain populations in LAO, driving a continuous and controlled transition from a four-variant rhombohedral crystal system to a two-variant orthorhombic crystal system at strain levels below $0.5\%$ or equivalently $\sim$ 0.7~GPa applied pressure. This transition is met by the complete removal of step terraces in the system along the strain axis, a fact of huge relevan\revision{ce} to communities working on LAO systems as an epitaxial substrate. The accompanying strain-induced lifting of the $E_g$ phonon degeneracy, confirmed by both Raman spectroscopy and first-principles calculations, provides a phonon-level fingerprint of the symmetry reduction underlying this transition, and establishes a quantitative spectroscopic measure of the domain state that could be exploited for non-contact characterisation in device geometries moving forwards. The continuous tunability of the domain wall angle raises the prospect of exploiting the LAO domain structure as an active degree of freedom in substrate-film systems: if the strain-driven domain redistribution can be shown to couple directly to the transport or magnetic properties of an overlying functional film, this would establish a practical route to \textit{in-situ} tunable oxide devices. More generally, the methodology demonstrated here, combining quantitative atomic force microscopy, three dimensional reciprocal space mapping, and Raman spectroscopy under \textit{in-situ} uniaxial strain, is directly transferable to the wider class of ferroelastic perovskite oxides, many of which share analogous twin domain structures and phase diagrams. Applied to systems where ferroelastic order is coupled to ferroelectric, magnetic, or orbital degrees of freedom, uniaxial strain could allow for the controlled manipulation of functional properties through domain engineering.
\clearpage


\subsection{Methods}
\subsubsection{Lanthanum Aluminate Sample And Implementation In Strain Cell}
Double sided polished LAO samples, with (001) orientation, from CRYSTAL GmbH, grown via the Czochralski method, are cut and further polished to obtain bars with the dimensions of $\sim$(5$\times$0.3$\times$0.2)~mm$^3$. These bars are glued with epoxy-resin onto a piezo-actuator driven titanium strain cells with force calibrated strain gauges (see Fig.~1d), which are readout by a Stanford Research Systems SR830 DSP lock-in amplifier. The methods of constructing and calibrating the home-made strain cells as well as the gluing with an ensured force transmission into the sample are well established and proven \cite{Pionteck_2025, Singh_2022, Singh_2023}. \revision{The sample is mounted in a way that it is strained along the (100) axis.} Knowing the cross section of the sample in combination with a continuous force measurement enables the determination of stress on the corresponding sample during the measurement. To investigate domain properties of the sample under strain a voltage is applied to the piezo-actuators, which is continuously either ramped up or down. In order to ensure at any point of straining a state which is sufficiently close to adiabatic, the ramping till the maximum strain takes typically two hours. Artifacts from e.g. hysteresis effects are suppressed by cycling the sample on the strain cell several times between the minimal and maximal strain before the actual measurements.
\subsubsection{Atomic Force Microscopy (AFM)}
The topography measurements are performed on an well established combination\cite{Roeper_2024, Singh_2022} of Park NX10 SFM and solid platinum tips 25Pt300B manufactured by Rocky Mountain Nanotechnology, LLC, in contact mode. Typical scanning parameters are (256$\times$256) Pixels with a 1.5~Hz scanning rate along the fast-scanning axis. The set-point used, of $\sim$700~nN, is sufficiently low to ensure that its influence on the dynamics and structures of interest is negligible. The trace and back-trace of the height and error signal are both recorded. If only the slope along one axis is of interest, the error signal is further analysed. Since the error signal is the deviation of the PID-loop the AFM-height controller is faced \cite{Park_Manual}, it is a measure of the local inclination of the sample surface, while a line alignment along the slow-scanning axis is intrinsically not required. In case the full two dimensional inclination has to be considered, the multi-dimensional Jacobian matrix is derived from the height signal. This distinction is done, because the error signal is slightly less noisy and avoids artifacts along the slow-scanning axis.     
\subsubsection{$\mu$-Raman Spectroscopy}
 The $\mu$-Raman spectroscopy measurements are acquired on a HORIBA LabRAM HR Evolution system, utilizing a 532~nm, 100~mW continuous wave laser in combination with a 100$\times$ magnification objective with a numerical aperture of 0.9 in a back-scattering geometry. The signal is detected by a Syncerity CCD camera (HORIBA Jobin Yvon GmbH) in combination with a 1800~lines/mm grating which leads to a spectral resolution of 0.013~nm or 0.48-0.68~cm$^{-1}$, respectively. To determine the peak splitting and softening with Raman spectroscopy, continuous measurements are performed, like the AFM measurements, while the strain is ramped up or down. Besides these comparatively fast measurements, which as a tradeoff cannot resolve the whole sample microscopically, maps with a resolution down to 0.5~$\mu$m are performed. Due to an integration time of 0.5~s per pixel with up to (252$\times$252) pixels, the ramping of the strain is stopped at certain levels to allow for longer measurement times. To determine the peak positions from the acquired Raman spectra the background including cosmic ray peaks are removed and a multi Voigt-fit\cite{Sundius_1973} is performed.   
 
\subsubsection{Density Functional Perturbation Theory (DFPT)}
For the numerical determination of the phonon dispersion and the corresponding spacial atomic vibrations by density functional perturbation theory\cite{Baroni_1987,Gonze_1995}, implemented in ABINIT \cite{Gonze2020, Romero2020} and Quantum Espresso\cite{Giannozzi_2009,Giannozzi_2017} are used. Scalar relativistic optimized norm-conserving Vanderbilt pseudo potentials \cite{Pseudo_Potentials} with Perdew-Burke-Ernzerhof exchange-correlation functionals are provided by Pseudo Dojo \cite{PseudoDojo}. The initial experimental crystal structure was selected from SpringerMaterials determined on a powder sample by an automatic neutron time-of-flight diffractometer \cite{sd_1814579}. This structure is then relaxed numerically.\\
The workflow to determine the atomic vibrations is set via AbiPy. After a converged relaxation \revision{(with a force convergence criterium of $10^{-5}~Ry\:\:a_0^{-1}$)}, the dynamical matrix of the system is calculated, based on a (4$\times$4$\times$4) k-point grid. The cutoff energy is set to 50~Ha.\\ 
In order to determine the strain induced shifts of the Raman peaks the energies at the $\Gamma$-point as well as the phonon dispersion for the maximum and minimum strain are calculated via Quantum Espresso\revision{, with a force convergence criterium of $10^{-4}~Ry\:\:a_0^{-1}$}. Linear Poisson effects are taken into account by the first principle compliance tensor. The compliance tensor is acquired by systematically analysing strain-stress tensor pairs of strain perturbed unit cells, while taking the symmetries of the system into account to avoid duplicated computational effort. The same computational parameters, as used beforehand, are used to model the strained unit cells.

\subsubsection{X-ray diffraction}
For the XRD measurement in the supplement, a Bruker Discover D8 with 1.6~kW Cu-Ka1 radiation with a wavelength of 1.54~\r{A} has been used with a parallel-beam geometry. After the Göbel mirror, the slits on the primary side are 0.2~mm and 6x2~mm. No additional optics are present on the secondary side. The detector is the Lynxeye XE-T. The tool has been controlled by the software DiffracMeasurementCenter and within this, the RSM has been performed with the WIZARD plug-in.

The three-dimensional reciprocal space mapping (3D-RSM) is performed using
a custom-built six-circle diffractometer. The diffractometer is equipped
with an Excillum liquid MetalJet source operating with a Ga/In alloy
target. A custom multilayer mirror X-ray optic selects the In K$\alpha$ doublet (0.512~\AA and 0.517~\AA) and focuses the beam to a spot size of 70x50 $\mu m^2$ with a beam divergence of 1.5~mrad. Diffracted intensity is recorded in reflection geometry using a Dectris PILATUS3 X CdTe 1M
hybrid photon-counting area detector. Data acquisition,
angular-to-reciprocal-space coordinate transformation, and subsequent
analysis are carried out using custom Python scripts built on the
open-source packages xrayutilities~\cite{Kriegner2013}.
\subsection{References}
\bibliographystyle{naturemag}
\bibliography{roeper_et_al_strain_1}

\subsection{Acknowledgments}
The authors gratefully acknowledge the computing time made available to them on the high-performance computer at the NHR Center of TU Dresden. This center is jointly supported by the Federal Ministry of Education and Research and the state governments participating in the NHR (\url{www.nhr-verein.de/unsere-partner}).
LM Eng, SC Kehr, J Wetzel acknowledge financial support by the Federal Ministery of Education and Research (BMBF, Germany, project grant nos. 05K190DB and 05K220DA) and by Deutsche Forschungsgemeinschaft (DFG, German Research Foundation) through the Würzburg-Dresden Clusters of Excellence ct.qmat – Complexity and Topology in Quantum Matter, and ctd.qmat – Complexity, Topology, and Dynamics in Quantum Matter (EXC 2147, project-id 390858490).
SD Seddon acknowledges Deutsche Forschungsgemeinschaft (DFG, German Research Foundation) through Project No. CRC1143 (ID 247310070) 
M~Roeper acknowledges ﬁnancial support by the Deutsche Forschungsgemeinschaft (DFG) through Project FOR5044 (ID: 426703838). 
R~Buschbeck and J Wetzel acknowledge financial support by the Deutsche Forschungsgemeinschaft (DFG, German Research Foundation) through Project No. CRC1415 (ID: 417590517).
I. Kiseleva contribution to this project was also co-funded by the European Union and co-financed from tax revenues on the basis of the budget adopted by the Saxon State Parliament.
Furthermore, the authors thank Thomas Gemming and Dina Bieberstein for assistance in
sample dicing. 
\subsection{Competing Interest Declaration}
The authors declare no competing interests.
\newpage
\subsection{Supplementary Information}
\subsubsection{Interpretation of Raman peak splitting}
The observed splitting of the $E_g$ modes originates from the fact that the under unstrained conditions degenerated $E_g$ modes are associate with energetically equivalent, but different lattice vibrations (only one vibration pattern is shown per degenerated pair in Fig.~4a), while the vibration of the $A_{1g}$ mode would, applying the same discrete symmetry operation (rotation around the (111) axis by $\pi$/2), end up in itself.

\subsubsection{Relaxation Of Lanthanum Aluminate Domains After Stress Removal}
Since the stress is ramped up to its maximum in a time regime of hours, the domain state of the domains within the sample is considered to be to first order quasi-static. In order to quantify higher order deviations AFM measurements at zero strain where carried out after cycling the stress of the sample several times (Fig.~S1). Each AFM scan is performed within a time of a few minutes. Relaxation effects, quantified from a higher order spacial deviation (domain measure Fig.~S1a), are measurable even several hours afterwards ($\tau$~=~4.8~h) and remaining sufficiently small to consider them as negligible for the performed measurements of the paper. However for higher cycle rates these effects becoming relevant. 
These spike domains present immediately after stress cycling, are also affecting in the XRD measurements (Fig.~S2).
This relaxation in zero strain allows furthermore the identification of the lower intensity shoulder peak in Fig.~2h, originating from the still active needle points remaining, and seemingly not linked to the domains themselves of the low strain regime. Constant AFM imaging of the measurement area indicated the motion of needle domains still moving throughout the sample (see supplement Fig.~S1), with their eventual stability broadly correlated with the removal of the shouldered XRD-peak seen in Fig.~S2g. These needle dynamics are subject to ongoing investigation.

\begin{figure}[tbh!]
    \centering
    \includegraphics[width=0.5\linewidth]{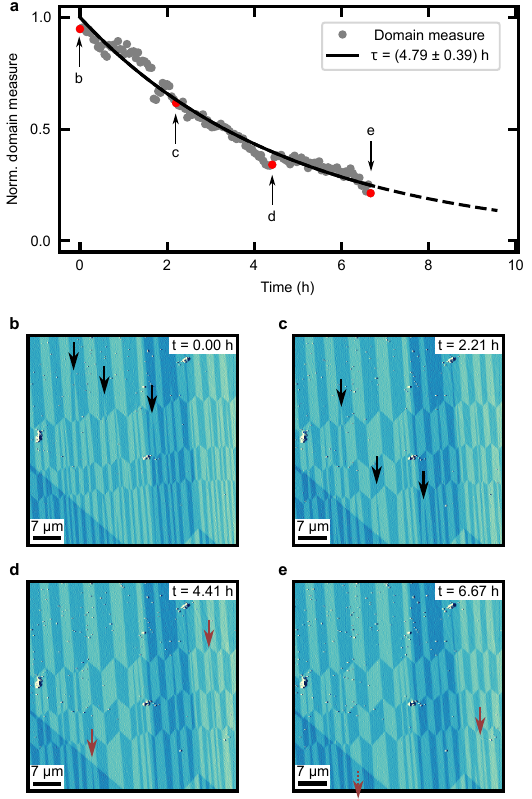}
    \caption{Domain relaxation captured by repeated AFM error signal measurements at different times. \textbf{a} Fit of the exponential decay of the domain activity after the external stress is released. \textbf{b} Measurement directly after the external stress is released. \textbf{e} Almost seven hours afterwards with intermediate measurements (\textbf{c}, \textbf{d}). Spike domains are still retracting from the upper part of the scan area towards the lower part under zero external stress. Selected spike domains retracting in (b-c) highlighted by black arrows and similarly for (d-e) indicated by red arrows.}
    \label{fig:placeholder}
\end{figure}

\begin{figure}
    \centering
    \includegraphics[width=0.9\linewidth]{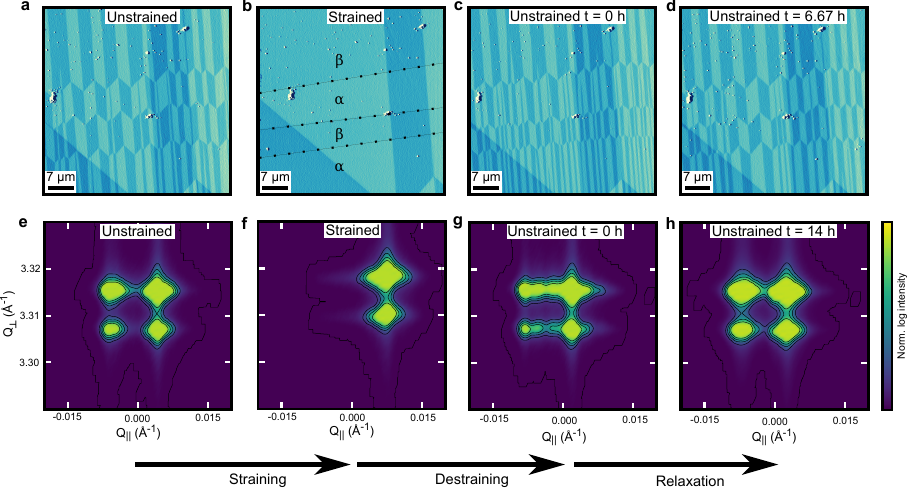}
    \caption{Strain cycling and relaxation qualified by the AFM error signal (\textbf{a}-\textbf{d}) and XRD intensity (\textbf{e}-\textbf{h)}. \textbf{a} AFM measurement of the initial unstrained sample with four domain types. \textbf{b} Sample at maximum strain with two faintly visible domain types $\alpha$ and $\beta$. \textbf{c} Destrained sample after several strain cycles with four types of domains, but compared to \textbf{a} with additional spike domains. \textbf{d} Unstrained sample after 6.67~h of relaxation. Almost all of the additional domains disappeared from the scanned area. \textbf{e-h} XRD measurement of an entirely relaxed, unstrained sample, under maximal strain, imediately at the point of zero applied strain, and after 14 hours of relaxation. Note the projection here includes vertical peak splitting due to the K$\alpha$ doublet, unlike the in-plane projection in the main text. Each reciprocal space map is analogue to the AFM image above it. 
    }
    \label{fig:xrd-relaxation}
\end{figure}
\subsubsection{Correlation of stress and strain}
Since the strain cell is equipped with sensors which are sensitive to the force applied to the sample, the stress can be, directly determined once knowing the samples cross section. In order to make statements with respect to the strain of the sample, XRD measurements of the sample mounted on the strain cell are performed, while ramping up and down the stress. The strain can be immediately calculated from the XRD based reciprocal lattice parameter $Q_x$. Since the dependence between stress and strain is sufficiently linear, a conversion factor is used whenever quantitative statements with respect to the strain are required. It has to be noted that while the statistical deviations remain small in the order of magnitude of 1\% or smaller, systematic deviations are accumulated from the conversion of the lock-in amplitude into a force, followed by the conversion into a stress value and finally into a strain value, resulting in a relative systematic deviation of not less than several 10\%. 

\end{document}